\documentclass[aps,prfluids]{revtex4-2}
\usepackage[table]{xcolor}
\usepackage{rotating}
\usepackage{amsmath}
\usepackage{bm}
\usepackage{array}
\usepackage{graphicx}
\usepackage{dcolumn}
\usepackage{amssymb}
\usepackage[english]{babel}
\usepackage{multirow}
\usepackage{lipsum}

\begin{document}

\title{Hydrodynamics constrain choanoflagellate collar geometry}

\author{Tasawar Iqbal}
\author{Catherine Penington}
\author{Christian Thomas}
 \affiliation{%
School of Mathematical and Physical Sciences \\ Macquarie University, NSW 2109, Australia
}
\author{Lyndon Koens}
\email{Corresponding Author: lyndon.koens@adelaide.edu.au}
\affiliation{
 Department of Mathematical and Physical Sciences\\
 University of Hull, UK \\
School of Mathematical Sciences,  Adelaide University, \\ Adelaide 5005, Australia
}
\begin{abstract}
As the closest living relatives of animals, choanoflagellates exhibit remarkable diversity. Even their microvilli collar, used to filter and capture food, varies significantly among species. This diversity suggests either strong environmental adaptation or an insensitivity to the collar geometry. Previous hydrodynamic studies have suggested that the pressure change across the collar is similar across species. In this study, we show that hydrodynamics imposes additional geometric constraints on the choanoflagellate collar. We create a simplified, reduced-order model that neglects finite collar length to investigate how the microvillus radius and the gap between microvilli influence the flow. Comparing with biological data reveals significant variation in the pressure drop between species. Additionally, a ridge emerges in the microvilli radius-gap phase space, along which both effective flux and power dissipation are maximised. Notably, several species cluster near the flux ridge but lie away from the power dissipation ridge. These observations suggest that choanoflagellate collars do not necessarily share a similar pressure drop. Instead, their geometry is influenced by the competing demands of maximising flux and minimising power costs. The broad variation observed among species is made possible by these ridge-like structures.
\end{abstract}
\maketitle

\section{Introduction}

Choanoflagellates are a class of aquatic microorganisms, found worldwide, and characterised by conical filter structures~\citep{leadbeater2015choanoflagellates}. These distinctive filters surround waving flagella, which draw food towards the structure for capture and consumption. The first experimental observations described the choanoflagellate collar as a mucus layer~\citep{leadbeater2015choanoflagellates}, but later imaging revealed it to be composed of interconnected arms called microvilli, whose geometry directly influences feeding. The thickness, spacing, and length of these villi can vary significantly between species. Further morphological variations between choanoflagellates include the formation of colonies, the creation of surrounding ``baskets" (called lorica), and whether species are freely suspended, swimming, or attached to surfaces (sessile).

Choanoflagellates play a key role in aquatic food webs, and their close relationship to animals makes them a popular model for studying the origins of multicellularity~\citep{kirkegaard2016motility,fung2023swimming}. Their aquatic nature and diversity have also attracted hydrodynamic interest~\cite{Fenchel1986, kirkegaard2016filter, kirkegaard2016motility, fung2023swimming, walther2018hydrodynamic, asadzadeh2019hydrodynamic, pettitt2002hydrodynamics, sorensen2021hydrodynamics, nguyen2019effects, Nguyen_Ross_Cortez_Fauci_Koehl_2023,  ayaz1999flow, Chen2025}. Early hydrodynamic models examined the permeability and porosity of an infinite array of posts~\citep{Fenchel1986,ayaz1999flow,tamada1957steady} to explore choanoflagellate feeding, revealing that the pressure drop across the filter was similar across different species. These findings led to the pressure drop hypothesis, which proposes that collar geometries maintain a roughly invariant pressure drop despite morphological variation. This interpretation was later reiterated by  \citet{pettitt2002hydrodynamics}, who developed a detailed model for three choanoflagellate species that broadly supported this conclusion, noting that while the qualitative trend predicted by the porous layer model was generally accurate, the magnitude was not. The hypothesis has since become influential in the biological literature, with the leading reference work on choanoflagellates~\citep{leadbeater2015choanoflagellates} identifying pressure drop as a key factor governing the filter geometry. Building upon these earlier works, organism-specific models have been developed to study the swimming and feeding behaviours of colonial choanoflagellates~\citep{kirkegaard2016filter, kirkegaard2016motility, fung2023swimming, Chen2025}, the influence of the lorica~\citep{walther2018hydrodynamic, asadzadeh2019hydrodynamic}, variations in cell morphology~\citep{nguyen2019effects}, and obstructions in the collar~\citep{sorensen2021hydrodynamics, Nguyen_Ross_Cortez_Fauci_Koehl_2023}.

This paper presents a simplified model of the choanoflagellate filter to revisit the evolutionary pressures involved in shaping its structure. The filter is represented as a porous medium, with the flagella modelled as an infinitely pulsating cylinder. Although the model does not account for finite collar length, it captures the leading-order behaviour near the filter, enabling investigation of how filter parameters influence the flow, pressure, and other quantities. Comparisons with biological data reveal that the maximum pressure drop across the filter can vary substantially between species. Nevertheless, many species cluster near a ridge in the “effective filter flux”, while lying further away from a similar ridge in the “effective power”. These ridges allow considerable biological variation, since changes in filter geometry along the ridge direction can have a small effect on the flux and power evolutionary pressure.

%%%%%%%%%%%%%%%%%%%%%%%%%%%%%%%%%%%%
%%%%%%%%%%%%%%%%%%%%%%%%%%%%%%%%%%%%
%%%%%%%%%%%%%%%%%%%%%%%%%%%%%%%%%%%%

\section{Methods}

\subsection{Theoretical model}

\begin{figure}
     \centering
     \includegraphics[width=0.9\textwidth]{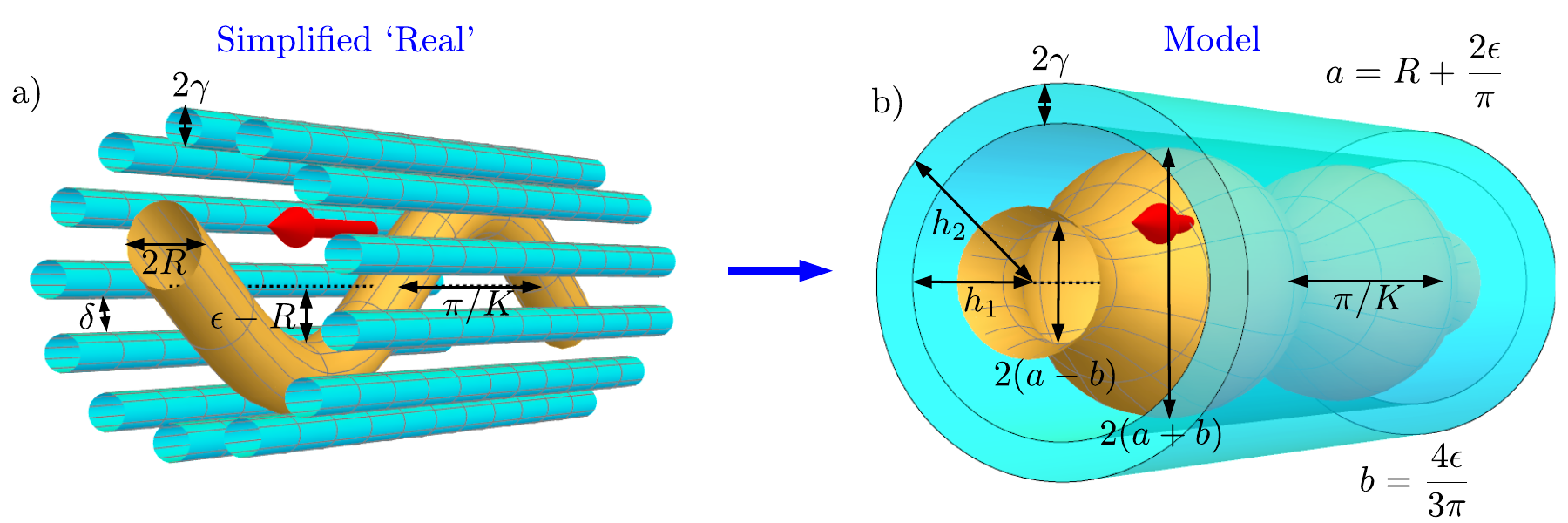}
     \caption[Model of the cylindrical Taylor swimming sheet surrounded by a ring of infinite cylinders]{a) An infinite waving filament surrounded by a ring of infinite cylinders, which represent the microvilli collar. b) An axisymmetric model of a), with a deforming cylinder and a Brinkman layer for the collar. The radius of the waving filament is $R$, the microvilli radius is $\gamma$, the gap between villi is $\delta$, the amplitude of the waving filament is $\epsilon$, and the wavelength is $2\pi/K$. Red arrows indicate the direction of the wave.}
     \label{fig:diagram}
 \end{figure}

The choanoflagellate filter consists of a long, central waving filament that drives the flow, surrounded by multiple microvilli that capture food. While finite microvilli typically sit in a conical orientation, we assume the conical tilt is small for simplicity. A section of the filter can therefore be approximated as an infinite waving filament surrounded by a ring of microvilli (see Fig.~\ref{fig:diagram}(a)), which we model as a ring of infinite cylinders. This infinite-filament reduction means the model does not capture the effects of microvilli length or the conical angle of the filter.

The infinite flagellum geometry remains complex. As a further simplification, we consider an axisymmetric version, representing an azimuthally averaged configuration that lies between the early porous-filter models \citep{Fenchel1986,ayaz1999flow,tamada1957steady} and the full-body numerical models \citep{pettitt2002hydrodynamics}. If a sinusoidally varying central flagella is rotated about its axis, the outermost surface of the resulting geometry satisfies 
\[\rho = R + \epsilon |\sin(z)|,\] 
where lengths are scaled by the flagellum wavenumber, $K$, velocities by the wave speed, $c$~\citep{taylor1951analysis,blake1971infinite,s7jz-7k79}, $\rho$ is the cylindrical radial coordinate, $R$ is the filament radius, $\epsilon$ is the wave amplitude, and $z$ the position in a frame moving with the wave~\cite{taylor1951analysis}. Similarly, under this rotational averaging, the ring of microvilli becomes an annular layer. Fluid must pass through this layer for filtering, so we treat it as a porous medium. The final reduced model, as depicted in Fig.\ref{fig:diagram}(b), therefore resembles Blake's cylindrical swimmer~\citep{blake1971infinite}, surrounded by an annular porous layer modelled by the Brinkman equations~\citep{ochoa1995momentum, mirbagheri2016helicobacter}. 

The axisymmetric collar geometry forms a three-fluid region model: a pulsating cylindrical filament, surrounded by a Brinkman fluid annulus, which allows the problem to be formulated in cylindrical coordinates $(\rho,\phi,z)$. In the waving cylinder reference frame, the governing equations in each region are 
\begin{subequations}\label{flow_eq}
\begin{align}
   \nabla^2 \boldsymbol{u}^{(1)} &= \nabla p^{(1)}, 
   \qquad 
   \nabla\cdot{\boldsymbol{u}^{(1)}} = 0,  
   \qquad 
   \mbox{for } R + \epsilon |\sin(z)|<\rho<h_1
   \tag{\theequation a,b} \\
   \nabla^2 \boldsymbol{u}^{(2)} - k^2  \boldsymbol{u}^{(2)} &= \nabla p^{(2)},   
   \qquad 
   \nabla\cdot{\boldsymbol{u}^{(2)}} = 0, 
   \qquad 
   \mbox{for } h_1<\rho<h_2
   \tag{\theequation c,d} \\
   \nabla^2 \boldsymbol{u}^{(3)} &= \nabla p^{(3)}, 
   \qquad 
   \nabla\cdot{\boldsymbol{u}^{(3)}} = 0,  
   \qquad 
   \mbox{for } h_2<\rho
   \tag{\theequation e,f} 
\end{align}
\end{subequations}
with the fluid stress
\begin{subequations}
\begin{align}
    \boldsymbol{\sigma}^{(1)}=- p^{(1)}\boldsymbol{I}+\nabla\boldsymbol{u}^{(1)}+(\nabla\boldsymbol{u}^{(1)})^T, \\
     \xi \boldsymbol{\sigma}^{(2)}=- p^{(2)}\boldsymbol{I}+\nabla\boldsymbol{u}^{(2)}+(\nabla\boldsymbol{u}^{(2)})^T, \\
      \boldsymbol{\sigma}^{(3)}=- p^{(3)}\boldsymbol{I}+\nabla\boldsymbol{u}^{(3)}+(\nabla\boldsymbol{u}^{(3)})^T, 
\end{align}
\end{subequations}
where $\boldsymbol{u}^{(i)}= u^{(i)} \hat{\boldsymbol{\rho}} + w^{(i)} \hat{\boldsymbol{z}}$ and $p^{(i)}$ denote the velocity and pressure in regions $i=1,2,3$, $(\cdot)^T$ is the transpose, and $k^2 = \xi \alpha^2/ K^2$~\cite{ochoa1995momentum}. Here, $\boldsymbol{u}^{(1)}$ represents the Newtonian flow between the waving cylinder and the start of the annulus, $\boldsymbol{u}^{(2)}$ represents a Brinkman flow in the annulus $h_1<\rho<h_2$, characterised by porosity, $\xi$ (fraction of the medium filled by fluid), and permeability, $1/\alpha^2$ (inverse of the layer resistance), and $\boldsymbol{u}^{(3)}$ is the Newtonian flow outside the Brinkman layer.

The porosity and permeability depend on the microvilli thickness, $\gamma$, and the distance between two villi centres, $\eta$. The gap between consecutive villi is $\delta=\eta - 2 \gamma$. Since porosity is defined as the fraction of the region occupied by fluid, it can be calculated from the area of the annulus and the area occupied by the microvilli. Each microvilli is assumed cylindrical, with cross-sectional area $\pi \gamma^2$. If the annulus starts at $\rho=h_1$, the number of microvilli around the cross section is roughly the circumference of a circle at the middle of the layer, $2 \pi (h_1+\gamma)$, divided by the distance between two microvilli centres, $\eta$. Hence, the area occupied by the microvilli in the layer is approximately $2 \pi^2 \gamma^2 (h_1+\gamma)/\eta$. The area of the annulus is $\pi (h_1+2\gamma)^2 - \pi h_1^2 = 4 \pi \gamma (h_1+\gamma)$. The porosity is then one minus the ratio of these two areas, giving
\begin{equation}
    \xi = 1 - \frac{\pi\gamma}{2\eta} .
\end{equation}
Similarly, the permeability, $1/\alpha^2$, is the mean flow divided by the pressure drop over the layer, and in our scaled coordinates becomes
\begin{equation}
    k^2 = \frac{D}{\eta}, 
\end{equation}
where $D$ is the drag per unit length on a microvilli. The drag, $D$, can be approximated using results for an infinite array of posts~\cite{ayaz1999flow,tamada1957steady,koens2023}, and asymptotically takes the form $D=8\pi/\Lambda$, where
\begin{equation}\label{eq:reverse v}
    \Lambda = 1-2\log{\tau} + \frac{1}{6}\tau^2 - \frac{1}{144}\tau^4 + \frac{1}{1080}\tau^6
    - \frac{53}{345600}\tau^8 + \frac{139}{5443200}\tau^{10} + O(\tau^{12}),
\end{equation}
for $\tau=2\pi\gamma/\eta$~\cite{ayaz1999flow,tamada1957steady}.

The surface of the pulsating cylinder can be approximated by $\rho= R + \epsilon |\sin(z)| \approx a+b \cos(2 z)$, where $a = R+2 \epsilon/\pi$ and $b=4 \epsilon / 3 \pi$, using the first two terms of the Fourier expansion of $|\sin(z)|$. Following Blake's cylindrical swimmer~\citep{blake1971infinite}, we impose the no-slip condition at this surface,
\begin{equation}
    \boldsymbol{u}^{(1)}|_{\rho=a+b\cos(2z)} = \hat{\boldsymbol{z}}-Q \hat{\boldsymbol{t}}, \label{boundary}
\end{equation}
where $\hat{\boldsymbol{t}} = \cos(\theta) \hat{\boldsymbol{z}} + \sin(\theta)\hat{\boldsymbol{\rho}}$, and
\begin{equation}\label{eq:scaled material mass A}
    Q=\frac{1}{2\pi}\int_{0}^{2\pi}{\sqrt{1+4 b^2\sin^2(2z)} \; \textrm{d}z}
\end{equation}
is the mass travelling back along the surface, and $\theta$ is the angle between the tangent to the wave and the $z$-axis \cite{sauzade2011taylor}. 

At the inner, $\rho=h_1$, and outer, $\rho=h_2=h_1+2 \gamma$, boundaries of the Brinkman annulus, continuity of velocity
\[\boldsymbol{u}^{(l)}_{\rho=h_{l}} =\boldsymbol{u}^{(l+1)}_{\rho=h_{l}}\] 
and normal stress 
\[\hat{\boldsymbol{\rho}} \cdot \boldsymbol{\sigma}^{(l)}_{\rho=h_{l}} =\hat{\boldsymbol{\rho}} \cdot \boldsymbol{\sigma}^{(l+1)}_{\rho=h_{l}}\] 
for $l=1$ and $2$, are enforced, representing well-motivated conditions derived from mean-field considerations~\citep{ochoa1995momentum}.

Observations of choanoflagellates indicate that the waving filament nearly touches the filter, suggesting that $h_1\approx a+b$~\cite{Mah}. Far from the filter, the flow approaches
\begin{equation}
    \lim_{\rho\to \infty} \boldsymbol{u}^{(3)} = U \hat{\boldsymbol{z}},
\end{equation}
where $U$ is the steady swimming speed of the cylinder. 

The axisymmetry allows the flow to be written in terms of the cylindrical streamfunction, $\psi$, defined by
\begin{subequations}\label{Streamfunction}
\begin{equation}
    u(\rho,z) = -\frac{1}{\rho}\frac{\partial{\psi}}{\partial{z}} 
    \quad \textrm{and} \quad
 w(\rho,z) = \frac{1}{\rho}\frac{\partial{\psi}}{\partial{\rho}}.\label{stream function} 
 \tag{\theequation a,b} 
\end{equation}
\end{subequations}
This representation automatically satisfies incompressibility and transforms the governing equations~\eqref{flow_eq} to
\begin{subequations}
\begin{equation}
    L_{-1}^2{\psi^{(i)}}-k \delta_{i2} L_{-1}{\psi^{(i)}}=0,
\end{equation}
where 
\begin{equation}
L_{-1}=\partial_{zz}+\partial_{\rho\rho}-(1/\rho)(\partial_{\rho}).
\end{equation}
\end{subequations}
General solutions to these equations are available in~\citep{blake1971infinite, ho2016swimming}.

A regular perturbation expansion is performed in the limit $b\ll 1$, following the method of \citet{taylor1951analysis}. The streamfunction, $\psi$, and swimming speed, $U$, are expanded as
\begin{subequations}\label{BC5}
    \begin{align}
     \psi^{(i)}(\rho,z)&=b \psi_1^{(i)}(\rho,z)+b^2\psi_2^{(i)}(\rho,z)+..., \label{eq:stream funtion perturbation} \\
     U&=b^2 U_2+b^3 U_3+... \, . \label{eq:far field velocity perturbation}   
    \end{align}
\end{subequations}
The structure of the above expansion is based on Taylor's original study \cite{taylor1951analysis}, in which he found that the flow starts at $O(b)$ whereas the swimming speed starts at $O(b^2)$.
Only the no-slip boundary condition, Eqn.~\eqref{boundary}, changes in this expansion, which becomes
\begin{subequations}\label{BC7}
\begin{align}
\label{eq:first order first stream boundary condition y component}
\frac{1}{\rho}\frac{\partial{\psi_{1}^{(1)}}}{\partial{\rho}}\bigg|_{\rho=a} &= 0, \\
\frac{\partial{\psi_{2}^{(1)}}}{\partial{\rho}}\bigg|_{\rho=a} +\cos(2 z)\frac{\partial}{\partial \rho} \left(\frac{1}{\rho}\frac{\partial \psi_1^{(1)}}{\partial \rho}\right)\bigg|_{\rho=a}
&=-a\cos(4 z), \label{eq:second order first stream boundary condition y component} \\
\label{eq:first order first stream boundary condition z component}
\frac{\partial{\psi_{1}^{(1)}}}{\partial{z}}\bigg|_{\rho=a}
 &= -2 a\sin(2 z), \\
 \frac{1}{\rho}\frac{\partial{\psi_{2}^{(1)}}}{\partial{z}}\bigg|_{\rho=a}
+\cos(2 z)\frac{\partial}{\partial \rho }\left(\frac{1}{\rho}\frac{\partial \psi_1^{(1)}}{\partial{z}}
\right)\bigg|_{\rho=a}
& = 0,\label{eq:second order first stream boundary condition z component}
\end{align}
\end{subequations}
where $a$ and $h_1$ are treated as $O(1)$. Mathematica~\cite{Mathematica} was used to solve the expanded equations to $O(b^2)$ to capture the leading-order effects of the swimming speed, $U$. Although the analytic solution is extensive, it reproduces a Taylor swimming sheet beneath a finite Brinkman layer~\cite{s7jz-7k79,blake1971infinite} for large $a$ (not shown). Presenting the solutions graphically captures the continuous behaviour and provides a level of clarity and completeness not possible with numerical simulations.

\begin{table}
\caption[Species Data]{Summary of species data on microvilli geometry. Unscaled data and references are listed in Tab.~\ref{tab-values} of appendix~\ref{AppendixA}.}\label{tab-1}
%\centering
%\footnotesize
\begin{ruledtabular}
\begin{tabular}{lcc}
Species &   Radius, $\gamma$ & Gap, $\delta$ \\
        \hline
        \textit{S. amphoridium}& $0.0175\pm0.0021$  & $0.15 \pm 0.12$ \\
        \textit{C. gracilis} & $0.0459 \pm0.0023$&$0.17 \pm 0.11$ \\
        \textit{S. diplocostata} & $0.0546\pm0.003$  &$0.20 \pm0.11$ \\ 
        \textit{D. grandis} & $ 0.051 \pm 0.004$  &$0.22 \pm 0.05$ \\ 
        \textit{M. sp.} & $0.0534$&$0.26\pm 0.05$ \\
        \textit{M. brevicollis} sessile & $0.0283$  &$0.175$ \\
        \textit{M. brevicollis} swim& $0.0283$  &$0.175$ \\
        \textit{S. rosetta} fast & $ 0.00828$  & $0.005$ \\
        \textit{S. rosetta} slow & $0.0313$  & $0.0322$ \\
        \textit{C. botrytis} & $0.030 \pm 0.009$  & $0.09 \pm 0.032$ \\
        \textit{C. flexa} flagella out & $0.0105\pm0.0021$  & $0.19\pm 0.04$ \\
        \textit{C. flexa} flagella in & $0.017\pm 0.006$  &$0.31\pm0.12$ \\
        \textit{M. ovata} & $0.0170$ & $0.10$
\end{tabular}
\end{ruledtabular}    
\end{table}

%%%%%%%%%%%%%%%%%%%%%%%%%%%%%%%%%%%%
%%%%%%%%%%%%%%%%%%%%%%%%%%%%%%%%%%%%
%%%%%%%%%%%%%%%%%%%%%%%%%%%%%%%%%%%%

\subsection{Biological data}

Data were collected from the literature on the flow and feeding mechanisms of choanoflagellates. A total of 13 species or forms were identified, and the relevant properties of the flagellum and microvilli were extracted or estimated from published figures. Table~\ref{tab-1} summarises the variation in the microvilli radius, $\gamma$, and the gap between microvilli, $\delta$ (see Tab.~\ref{tab-values} in Appendix~\ref{AppendixA}). The biological data suggest an average flagella radius of $R=0.069 \pm 0.0027$ and a wave amplitude of $\epsilon = 1.13 \pm 0.11$, corresponding to $b \approx 0.48$. These values of $R$ and $\epsilon$ are used throughout unless otherwise specified. Based on previous studies \cite{Mah, nielsen2017hydrodynamics}, we also take $h_1=b$ throughout.

%%%%%%%%%%%%%%%%%%%%%%%%%%%%%%%%%%%%
%%%%%%%%%%%%%%%%%%%%%%%%%%%%%%%%%%%%
%%%%%%%%%%%%%%%%%%%%%%%%%%%%%%%%%%%%

\section{Results}

The simplified model captures leading-order hydrodynamic behaviour, enabling direct comparison with the biological data. The analysis considers the maximum pressure drop, $\Delta p$, the maximum fluid velocity entering the filter, $u_{\max}$, the velocity along the filter, $U$, and the surface area of the microvilli, $S$, defined as 
\begin{subequations}
\begin{eqnarray}
    \Delta p &=& \max p^{(2)}(h_1+2\gamma,z)-p^{(2)}(h_1,z), \\
    u_{\max} &=& \max u^{(3)}(h_1+2\gamma,z), \\
    S&=& 4 \pi^2 \gamma (h_1+\gamma)/\eta.
\end{eqnarray}
\end{subequations}
Geometrically, the surface area of the villi, $S$, is the product of the cylinder perimeter, $2 \pi \gamma$, and the total number of cylinders, $2\pi (h_1+\gamma)/\eta$. In this model, the average velocity into the filter is zero as a consequence of the infinite periodic formulation, while the velocity along the filter equals the far-field swimming speed. In real filter feeders, however, a flow enters the filter and exits out of the collar mouth. Despite this difference, the reduced infinite model provides qualitative insight into filter behaviour, similar to how Taylor’s swimming sheet captures the behaviour of finite swimmers (even though real swimmers generate a net flux that the infinite model cannot).

\begin{figure}
\centering
\includegraphics[width=0.8\textwidth]{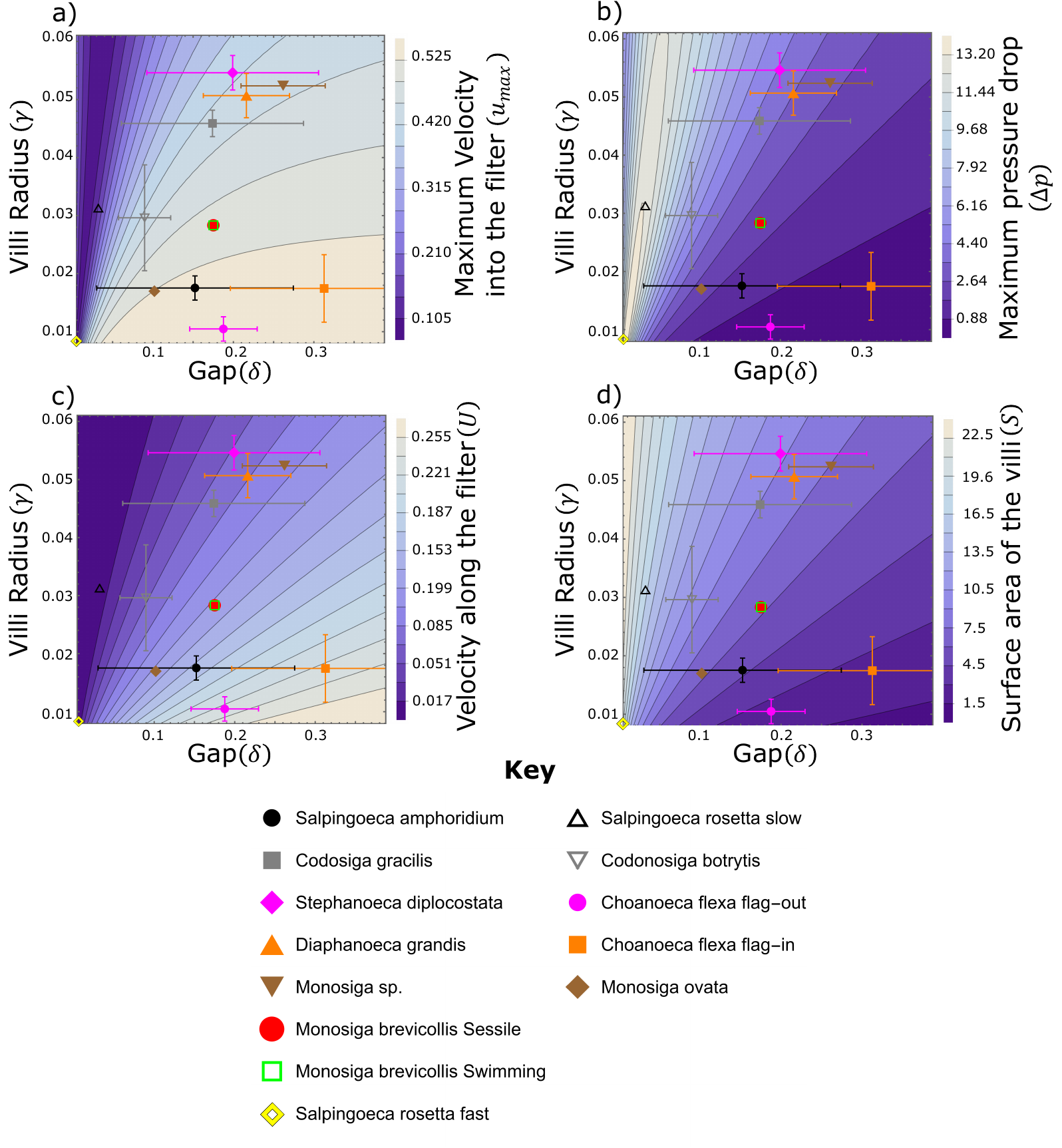}
\caption{a) The maximum velocity into the Brinkman layer, $u_{\max}$, b) the maximum pressure drop, $\Delta p$, c) the velocity along the filter, $U$, and d) the surface area of the filter arms, $S$, as contours within the plane defined by the microvilli radius, $\gamma$, and the gap between microvilli, $\delta$. Here, $R=0.069$ and $\epsilon =1.13$. Data points with error bars represent measurements of different species of choanoflagellates using the biological data from Tab.~\ref{tab-1}. (\textit{M. brevicollis} sessile and \textit{M. brevicollis} swim have identical villi radius and gaps, as reported in Tab.~\ref{tab-1}. \textit{S. rosetta} fast located in bottom left-hand corner.)}
\label{fig:raw}
\end{figure}

Reducing the microvilli radius, $\gamma$, or increasing the gap between villi, $\delta$, enhances both the maximum fluid velocity entering the filter, $u_{\max}$, and the swimming speed, $U$ (see Figs.~\ref{fig:raw}(a) and~\ref{fig:raw}(c)). Simultaneously, these changes reduce the surface area of the filter, $S$, and typically reduce the maximum pressure drop, $\Delta p$ (see Figs.~\ref{fig:raw}(b) and~\ref{fig:raw}(d)). These trends are equivalent to removing obstacles in the flow. The apparent maximum in $\Delta p$ at small $\delta$ reflects the breakdown of the asymptotic approximation for $\Lambda$ (recall Eq.~\eqref{eq:reverse v}). The choanoflagellate data are distributed across 
%these gradients 
the parameter space in the $(\delta,\gamma)$-plane. Notably, the maximum pressure drop varies by an order of magnitude across the range shown in Fig.~\ref{fig:raw}(b), contradicting the hypothesis that choanoflagellate collars exhibit similar pressure drops. However, sessile, non-colonial choanoflagellates (\textit{Salpingoeca amphoridium, Codosiga gracilis, Stephanoeca diplocostata}) display comparatively similar maximum pressure drops, consistent with previous observations. 

\begin{figure}
     \centering
     \includegraphics[width=0.8\textwidth]{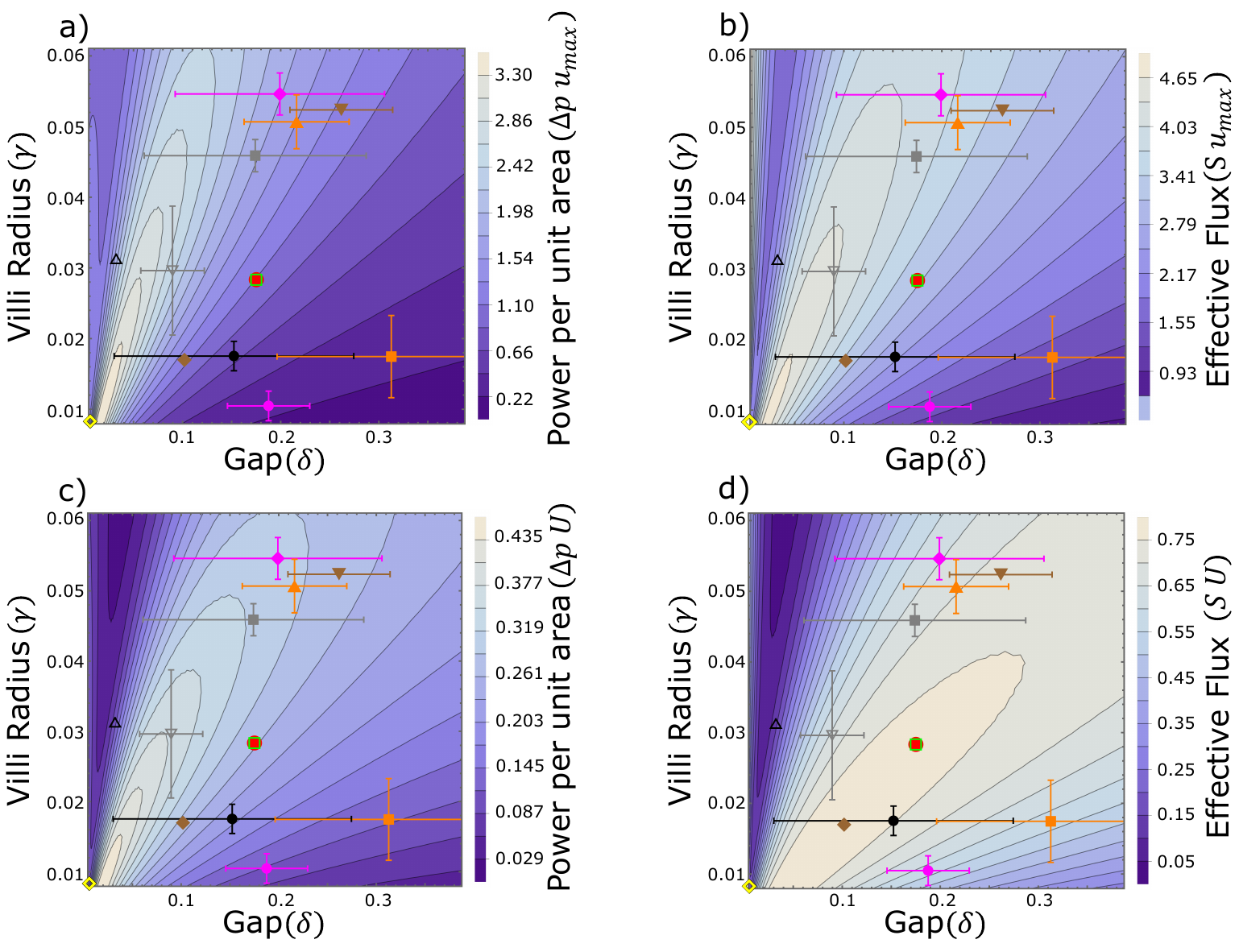}
     \caption{a) and c) The effective power per unit area, and b) and d) the effective flux. Here, $R=0.069$ and $\epsilon =1.13$. a) $\Delta p u_{\max}$ and b) $S u_{\max}$ represent the power and flux using the maximum flux into the filter, $u_{\max}$, while c) $\Delta p U$ and d) $S U$ are the power and flux using the velocity along the filter, $U$. Biological data symbols are as given in the legend of Fig.~\ref{fig:raw}. (\textit{M. brevicollis} sessile and \textit{M. brevicollis} swim have identical scales, as reported in Tab.~\ref{tab-1}. \textit{S. rosetta} fast located in bottom left-hand corner.)}
     \label{fig:PF}
\end{figure}

\begin{figure}
     \centering
     \includegraphics[width=0.8\textwidth]{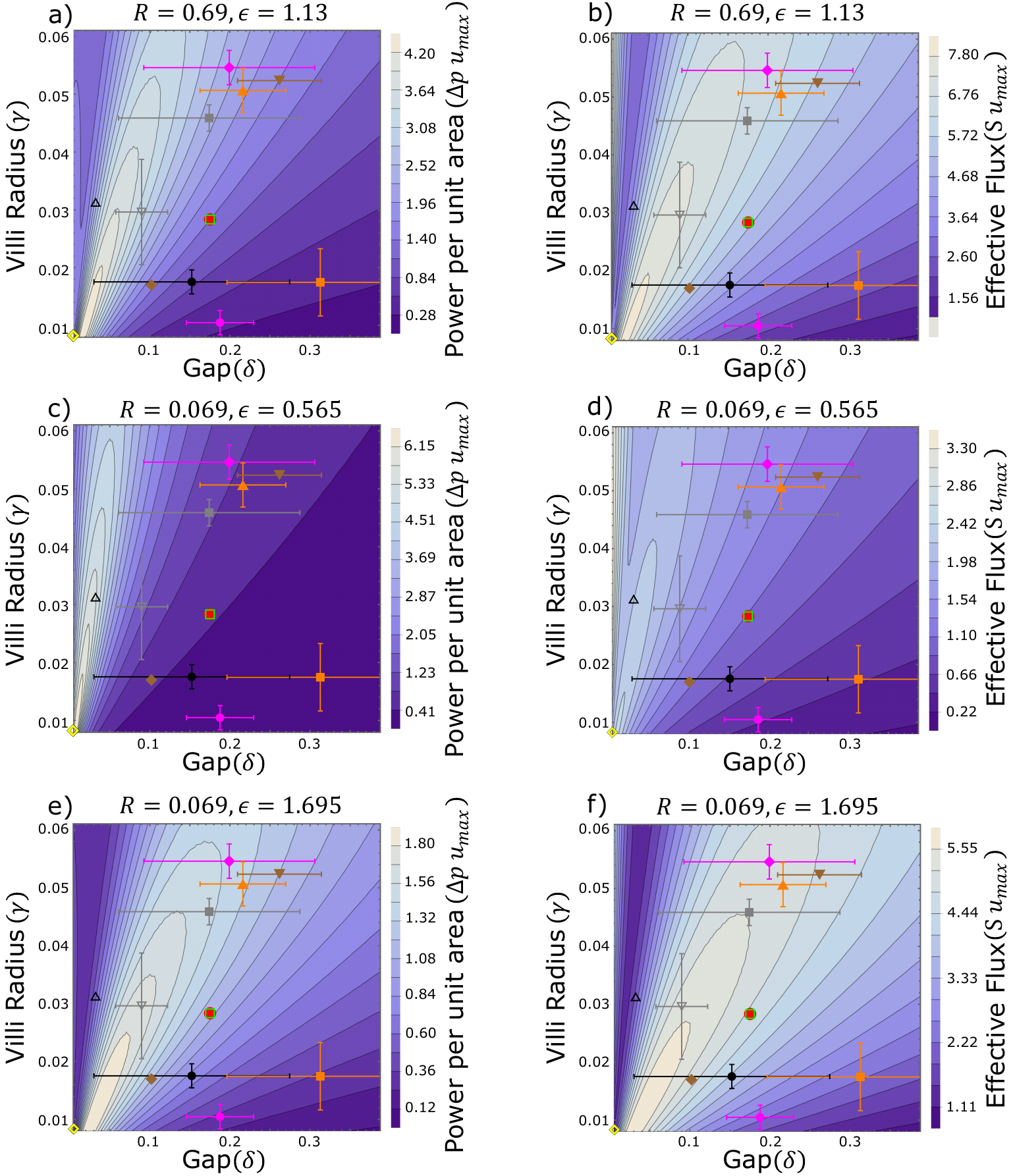}
     \caption{Effect of changing the flagellum radius, $R$, and wave amplitude, $\epsilon$, on the (a, c, e) effective power per unit area, and (b, d, f) effective flux. (a) and (b) show results for $R=0.69$ and $\epsilon =1.13$. (c) and (d) for $R=0.069$ and $\epsilon =0.565$. (e) and (f) for $R=0.069$ and $\epsilon =1.695$. All previous figures use $R=0.069$ and $\epsilon =1.13$. Biological data symbols are as given in the legend of Fig.~\ref{fig:raw}. (\textit{M. brevicollis} sessile and \textit{M. brevicollis} swim have identical scales, as reported in Tab.~\ref{tab-1}. \textit{S. rosetta} fast located in the bottom left-hand corner.)}
     \label{fig:Reps}
\end{figure}

The effective power per area (the product of $\Delta p$ and $u_{\max}$ or $U$) and the effective flux (the product of $S$ and $u_{\max}$ or $U$) form ridges in the $(\delta,\gamma)$-plane (see Fig.~\ref{fig:PF}). These ridges originate near the origin and extend approximately linearly. The flux ridge is less steep and occurs along the decreasing edge of the power ridge, in the direction of larger $\delta$ and smaller $\gamma$. Moreover, the power and flux computed using the maximum velocity, $u_{\max}$, create steeper ridges compared to those obtained using the velocity along the filter, $U$ (see Fig.~\ref{fig:PF}), with the latter forming a broader flux ridge. Reducing the distance to the Brinkman layer, $h_1$, increases the ridge slope, whereas varying other parameters (e.g., $h_2$, $k$, $\alpha$, $b$, and $a$) leaves the ridge shape qualitatively unchanged. Changing the flagellum radius, $R$ (see Figs.~\ref{fig:Reps}(a) and~\ref{fig:Reps}(b)), produces negligible qualitative changes, while decreasing the wave amplitude, $\epsilon$, steepens the ridges (see Figs.~\ref{fig:Reps}(c) and~\ref{fig:Reps}(d)), and increasing $\epsilon$ flattens the ridges (see Fig.~\ref{fig:Reps}(e) and (f)). Since solutions based on the maximum velocity entering the filter, $u_{\max}$, provide a more accurate geometric representation of both power dissipation and flux across the filter, the subsequent discussion focuses on these results.

%%%%%%%%%%%%%%%%%%%%%%%%%%%%%%%%%%%%
%%%%%%%%%%%%%%%%%%%%%%%%%%%%%%%%%%%%
%%%%%%%%%%%%%%%%%%%%%%%%%%%%%%%%%%%%

\section{Discussion}

For the parameter values $R=0.069$ and $\epsilon=1.13$, the biological data cluster around the ridge in the plot of the effective flux (see Fig.~\ref{fig:PF}(b)). Since flux, defined as the product of the velocity within the layer and the surface area of the villi, is related to filter-feeding rates, choanoflagellates may try to maximise this quantity. The clustering of choanoflagellates near the ridge is consistent with the hypothesis that they experience evolutionary pressure associated with maximising flux. However, the remaining variability suggests a trade-off with other ecological factors, such as resource availability, competition, and predation. 

Power usage may be one of these additional factors. The offset between the power and flux ridges permits higher flux at lower power cost. Interestingly, most of the choanoflagellate data lie on the side of the flux ridge farther from the power ridge, which is consistent with the hypothesis that flux can be increased while minimising power expenditure. 

The colony species \textit{Choanoeca flexa} and \textit{Salpingoeca rosetta} are located farthest from both the flux and power ridges. Notably, \textit{S. rosetta} lies on the opposite side of the power ridge from all other data points. The filter configurations on this side of the ridge feature smaller gaps between the villi and resemble sponge choanocyte filter configurations \cite{Mah}. The location of these two colony species results in significantly lower effective power costs compared to other species. This suggests that, in a colony, minimising the power expenditure of each individual may be advantageous, possibly due to the need for food distribution or to enhance collective flow. 

The position of \textit{Salpingoeca rosetta} relative to the ridges may reflect its smaller wave amplitude, $\epsilon$, which affects the slope of the ridges.
\textit{S. rosetta} tends to have a smaller wave amplitude in the collar than many of the other species considered (see Tab.~\ref{tab-values}), with $\epsilon \sim 0.1$-$0.5$. Hence, the corresponding ridges are steeper in the $(\delta,\gamma)$-plane, as shown in Figs.~\ref{fig:Reps}(c) and~\ref{fig:Reps}(d) for $\epsilon=0.565$, meaning \textit{S. rosetta} may also lie close to a maximising ridge. In this region, however, the approximations used in the permeability (see Eq.~\eqref{eq:reverse v}), are the least accurate. Notably, the same argument does not apply to \textit{Choanoeca flexa}, which also has a smaller $\epsilon$ value in the collected data.

%%%%%%%%%%%%%%%%%%%%%%%%%%%%%%%%%%%%
%%%%%%%%%%%%%%%%%%%%%%%%%%%%%%%%%%%%
%%%%%%%%%%%%%%%%%%%%%%%%%%%%%%%%%%%%

\section{Conclusion}

The significant variation in filter configurations among choanoflagellates appears to be hydrodynamically influenced by the need to maximise flux while minimising power dissipation. This observation contrasts with the prevailing hydrodynamic hypothesis, which suggests that filters maintain similar pressure drops across different species~\cite{leadbeater2015choanoflagellates}. Our simplified model, which neglects finite collar length effects, exhibits considerable variation in pressure among all species but similar drops for the original sessile non-colony species investigated. Notably, the flux and power display ridge extrema in the microvilli radius and microvilli gap phase space. Most of the biological data cluster near the flux ridge and away from the power ridge, suggesting that maximising flux plays a key role in shaping the collar geometry. The large variations in collar geometry may be a consequence of this ridge as changes along the ridge direction do not significantly change the flux and power. A specific species geometry likely depends on other biological or environmental factors not captured by the reduced-order model presented herein. The identification of these ridges provides insight into the relationships between choanoflagellate geometries and may be used to test how specific species respond to other evolutionary pressures.

%%%%%%%%%%%%%%%%%%%%%%%%%%%%%%%%%%%%
%%%%%%%%%%%%%%%%%%%%%%%%%%%%%%%%%%%%
%%%%%%%%%%%%%%%%%%%%%%%%%%%%%%%%%%%%

\section*{Acknowledgments} 

T.I. is funded by HEC Overseas Ph.D. Scholarship Batch II.

%%%%%%%%%%%%%%%%%%%%%%%%%%%%%%%%%%%%
%%%%%%%%%%%%%%%%%%%%%%%%%%%%%%%%%%%%
%%%%%%%%%%%%%%%%%%%%%%%%%%%%%%%%%%%%

\section*{Declaration of interest}

The authors report no conflict of interest.

%%%%%%%%%%%%%%%%%%%%%%%%%%%%%%%%%%%%
%%%%%%%%%%%%%%%%%%%%%%%%%%%%%%%%%%%%
%%%%%%%%%%%%%%%%%%%%%%%%%%%%%%%%%%%%

\section*{Data availability}

The data are not publicly available but are available from the authors upon reasonable request.

%%%%%%%%%%%%%%%%%%%%%%%%%%%%%%%%%%%%
%%%%%%%%%%%%%%%%%%%%%%%%%%%%%%%%%%%%
%%%%%%%%%%%%%%%%%%%%%%%%%%%%%%%%%%%%

\appendix

\section{Biological characteristics and microvilli geometry}\label{AppendixA}

Biological data were compiled from 11 references examining the swimming and flow behaviours of various choanoflagellate species (see Tab.~\ref{tab-values}). For each species, we recorded whether the organism was sessile or freely swimming, along with measurements of cell radius, flagella length, wave frequency, wavelength, microvilli radius, separation between microvilli, and collar villi length. When available, wave amplitude and swimming or flow speed were also recorded. All values were taken from the original sources, with the wavelength of \textit{Monosiga ovata} obtained from Ref.~\citep{nielsen2017hydrodynamics} due to the lack of an alternative source.
 
\begin{table}\caption[Species Data]{Biological data summarising species characteristics and microvilli geometry. All lengths are in $\mu$m, frequency in Hz, and speed in $\mu$ms$^{-1}$. The listed species are: 1 - \textit{Salpingoeca amphoridium}, 2 - \textit{Codosiga gracilis}, 3 - \textit{Stephanoeca diplocostata}, 4 - \textit{Diaphanoeca grandis}, 5 - \textit{Monosiga sp.}, 6 - \textit{Monosiga brevicollis} Sessile, 7 - \textit{Monosiga brevicollis} swimming, 8 - \textit{Salpingoeca rosetta} fast, 9 -  \textit{Salpingoeca rosetta} slow, 10 - \textit{Codonosiga botrytis}, 11 - \textit{Choanoeca flexa} flagella out, 12 - \textit{Choanoeca flexa} flagella in, 13 - \textit{Monosiga ovata}.}
\label{tab-values}
\begin{ruledtabular}
    \begin{tabular}{r c c c c c c c}
        Species & 1 &  2  &  3 & 4 & 5 & 6 & 7  \\ 
        \hline 
        Sessile & Y & Y & Y & N & N& Y &N  \\ 
        Cell radius & $2.3\pm0.7$ & $1.84 \pm 0.16$ & $1.81 \pm 0.63$ & $2.65\pm 0.15$ & $3.25\pm0.25$ &$2$ &$2$ \\
        Flagella length & $20.7\pm2.9$ & $8.3\pm 1.83$ &$8.58\pm1.95$ & $11.7\pm 1.5$ & $5.5\pm0.5$ & $11\pm0.5$&$11.2\pm0.4$ \\  
        Wave frequency & $17 \pm 0.31$ & $10 \pm 0.33$& $10$ & $7.3\pm2.6$ & $32.5\pm2.5$ & $49.9\pm2.4$ & $68\pm2.7$ \\ 
        Wave length & $17.92 \pm 1.15$ & $10.27\pm0.51$& $8.63\pm0.47$& $9.3\pm0.7$&$6$ &$12.2$ &$12.2$ \\
        Wave amplitude & -& -&-& $3.4\pm0.6$ & $1$& $2.4$ &$2.4$  \\ 
        Swimming/flow speed &$26 \pm 2.36$ & $20.5 \pm 2.11$ & $14\pm1.28$& - & $30$& - & - \\
        Villi radius & $0.05 \pm 0.005$ & $0.075$& $0.075$ &$0.75$ & $0.05$ & $0.055$ & $0.055$ \\
        Separation between villi & $0.53 \pm 0.35$ & $0.43 \pm 0.18$ & $0.42 \pm 0.14$ & $0.47 \pm 0.07$ & $0.35\pm0.05$ & $0.45$ & $0.45$ \\
        Villi length & $8.1 \pm 2.5$ & $4.5 \pm 1.5$& $4.9\pm1.1$ & $5$&$1.6\pm0.4$ & $7.3 \pm 0.5$ & $2.1\pm0.2$ \\ \hline
        References & \cite{pettitt2002hydrodynamics} & \cite{pettitt2002hydrodynamics} & \cite{pettitt2002hydrodynamics}& \cite{nielsen2017hydrodynamics}& \cite{fenchel1982ecology} & \cite{nielsen2017hydrodynamics,Mah}& \cite{nielsen2017hydrodynamics,Mah} \\  
 \\
 \\
 %\\
 Species & 8 & 9 & 10 & 11 & 12 & 13 \\ 
 \hline 
        Sessile & N &N & Y& N & N & Y \\ 
        Cell radius & $1.5$ & $1.85$ & $5$& $3.3$ & $3.3$ &$ 1.8 \pm 0.23$\\
        Flagella length & $ 47.8$ & $15.4$ & $27.5 \pm 2.5$ & $23\pm4$ &$26\pm5$ & $6$\\  
        Wave frequency &$24.3$& $24.3$ & $30$& $43\pm8$&$45\pm4$& $14.4\pm1.9$\\ 
        Wave length & $34.90$ &$9.23998$ & $17.5\pm 2.5$ & $15\pm3$&$9\pm3$ & $18.5$\\
        Wave amplitude & $0.61$& $0.71$ & - & $2.2\pm0.7$& $2.4\pm0.6$ & -\\ 
        Swimming/flow speed &$64$&$26$ & -& $67\pm20$ & $93\pm24$ & $9.3 \pm 5.7$\\
        Villi radius & $0.046$ & $0.046$ & $0.0825 \pm 0.045 $& $0.025$& $0.025$ & $0.05$ \\
        Separation between villi &$0.119$ &$0.139$ & $0.3325$& $0.498$ & $0.498$ & $0.4$ \\
        Villi length &$ 0.6$ & $5.8$& $ 9 \pm 1$& $10$ & $10$ & $4.5 \pm 1.5$ \\\hline
        References & \cite{Nguyen2019} & \cite{Nguyen2019}& \cite{Sleigh64,Fjerdingstad1961} & \cite{fung2023swimming,Brunet2019} & \cite{fung2023swimming,Brunet2019} & \cite{boenigk2000comparative,karpov1997cell,Leadbeater_1972,nielsen2017hydrodynamics} \\   
    \end{tabular}   
\end{ruledtabular}
\end{table}

\bibliography{references}
\end{document}